\documentclass [12pt] {report}
\usepackage{amssymb}
\newcommand {\D}[2] {\displaystyle\frac{\partial{#1}}{\partial{#2}}}

\newcommand {\al} {\alpha}

\newcommand {\ga} {\gamma}

\newcommand {\Si} {\Sigma}

\newcommand {\De} {\Delta}

\newcommand {\fr} {\displaystyle\frac}

\newcommand {\be} {\begin{equation}}
\newcommand {\ee} {\end{equation}}
\newcommand {\ba} {\begin{array}}
\newcommand {\ea} {\end{array}}
\newcommand {\bp} {\begin{picture}}
\newcommand {\ep} {\end{picture}}
\newcommand {\bc} {\begin{center}}
\newcommand {\ec} {\end{center}}
\newcommand {\bt} {\begin{tabular}}
\newcommand {\et} {\end{tabular}}
\newcommand {\lf} {\left}
\newcommand {\rg} {\right}

\newcommand {\bls} {\baselineskip}

\newcommand {\cH} {{\cal H}}

\newcommand {\ses} {\medskip}

\newcommand {\cP} {{\cal P}}
\newcommand {\bibit} {\bibitem}
\newcommand {\nin} {\noindent}

\def\2#1#2#3{{#1}_{#2}\hspace{0pt}^{#3}}
\def\3#1#2#3#4{{#1}_{#2}\hspace{0pt}^{#3}\hspace{0pt}_{#4}}
\newcounter{sctn}
\begin{document}

\begin {titlepage}

\vspace{0.1in}

\begin{center}
{\Large
Finslerian Post-Lorentzian Kinematic Transformations
}\\
\end{center}

\begin{center}
{\Large
in Anisotropic-Space Case
}\\
\end{center}

\vspace{0.3in}

\begin{center}

\vspace{.15in}
{\large G.S. Asanov\\}
\vspace{.25in}
{\it Division of Theoretical Physics, Moscow State University\\
117234 Moscow, Russia\\

 asanov@newmail.ru

}
\vspace{.05in}

\end{center}

\begin{abstract}

The Finslerian post-Lorentzian
kinematic transformations can explicitly be obtained under
uni-directional
breakdown of
spatial isotropy,
 provided that  the requirement that
the relativistic
unit hypersurface
 (indicatrix or
mass shell) be a space of
constant negative curvature is still
fulfilled.
The method consists in evaluating respective Finslerian tetrads and then
treating them as
the bases of inertial reference frames.
The Transport Synchronization has rigorously
been proven, which opens up the ways proper to favour the concept of
one-way light velocity.
Transition to the Hamiltonian treatment is
straightforward, so that
the
Finslerian
transformation laws for momenta and frequences,
as well as due Finslerian corrections to Doppler effect,
become
 clear.
An important common feature of the ordinary
pseudo-Euclidean theory of special relativity
and of the Finslerian relativistic approach under study
is that
they both endeavour to establish a universal prescription
for applying the theory to systems in differing states of motion.
\ses
\ses

\nin
\ses
\ses

\nin
Abbreviations: SR, FG, RS, and RF will be used for
special relativity, Finsler geometry,
reference system,
and
(inertial) reference frame,
 respectively.
\ses
\ses

\end{abstract}

\end{titlepage}

\vskip 1cm

{\nin\bf 1. Introduction}
\medskip

\nin
Kinematically,
the FG-generalization of the pseudo-Riemannian relativistic
theory of space-time may be seen to consist in that the metric tensor
is admitted to depend on motion velocity of local observer.
Involved SR-ideas will be discussed and contrasted
with their possible FG-extensions in Section 2.

When going beyond square-root metric, it occurs
constructive to keep the fundamental principle:
\ses

\nin
${\cP}_1$: {\it Indicatrix (mass shell)
is a space of constant negative curvature}.
\ses

\nin
The associated indicatrix extends ordinary relativistic
hyperboloid on a tiny deformation
estimated by a small characteristic parameter $g$.

The pseudo-Euclidean metric function  has long been a
convenient theoretical
cornestone of relativistic physics
and
our intuition is well developed
for such a metric.
Nevertheless,
the empirical constraints on the explicated Lorentz invariance,
as well as on the very metric,
are difficult to express in an exact way:
\ses

\nin
{\it ``...in practice, however, it is often impossible to
disentangle Lorentz invariance from other
theoretical issues and experimental complications''}
(M.P.Haugan and C.M.Will [1], p. 69).
\ses

\nin
That is to say, the degree of agreement can be estimated but outwardly,
on the basis of
a particular successful extension enabling one to
think of the agreement in terms of upper
bounds of the characteristic parameters of the extension.

In this vein, we  start with  stipulation of
a local preferred RF, to be denoted as $\Si$,
in which Lorentz invariance
is expected to breakdown because of a physically-distinguished spatial
direction.
To get a systematic Finslerian spatially-anisotropic SR-framework,
we  are to substitute a relevant FMF with the ordinary pseudo-Euclidean function
at the starting point of relativity theory.
To this end we act in several proper steps.

First, we set forth the concept
\ses

\nin
{\bf Finslerian Geometrically-Distinguished Direction},
to be denoted as
$\cal F\cal D$-direction.
\ses

\nin
Namely, let  a small
space-time
local region $U$
be chosen and endowed with
ordinary square-root relativistic metric
$S(T,X,Y,Z)
=\sqrt{T^2-X^2-Y^2-Z^2}$.
We interpret the coordinate set
$\{T,X,Y,Z\}$
as being related to a ``rest frame" $\Si$ meaningful over $U$.
Let
$\stackrel{\to}D$
be a vector in $U$.
We say that
$\stackrel{\to}D$
assigns a
$
\cal F\cal D$-direction
in $U$ if the account of such direction deforms
the metric $S$:
\be
S(T,X,Y,Z)
\stackrel{\cal F\cal D}{\Longrightarrow}
\Phi(
\stackrel{\to}D;T,X,Y,Z)
\ee
to get a Finslerian metric function
$\Phi$.

Second,
we  fix the case when
$\stackrel{\to}D$
is {\it of spatial nature}
and imply the correspondence principle
\be
\Phi(
\stackrel{\to}{D}=0;T,X,Y,Z)
=
S(T,X,Y,Z)
\ee
to hold
safely.

Afterward,  the rigorous isotropy is assumed to take place
around
the $\cal F\cal D$-direction.
That is, chosing the $R^1$-axis to point in the
the $\cal F\cal D$-direction,
and re-labelling the coordinates whenever convenient:
$\{T,X,Y,Z\}\to
\{R^0,R^1,R^2,R^3\}$,
we ascribe to the dependence the representation
\be
\Phi(
\stackrel{\to}D;R^0,R^1,R^2,R^3)
=
F\lf(g;R^1,\sqrt{(R^0)^2-(R^2)^2-(R^3)^2}\rg),
\ee
where $g$ is a small parameter for due experimental estimations.

Finally, we follow the Principle
${{\cal P}_1}$ formulated above.

This way proves to fix the FMF $F$ and the associated Finslerian
Hamiltonian function
$H$
as follows.

\be
F(g;R)=\lf|R^1-g_-q\rg|^{G_+/2}\lf|R^1-g_+q\rg|^{-G_-/2}
\ee
and
\be
H(g;P)=\lf|P_1-g_+\hat q\rg|^{-G_-/2}
\lf|P_1-g_-\hat q\rg|^{G_+/2},
\ee
where
\be
q=q(R)=
\sqrt{|(R^0)^2-(R^2)^2-(R^3)^2|}
\ee
and
\be
\hat q=
\hat q({P})=
\sqrt{|(P_0)^2-(P_2)^2-(P_3)^2|};
\ee
we use
the notation
$
G= g/h,\,
 h=(1+\fr14g^2)^{1/2},\,
g_+=-\fr12g+h,\,  g_-=-\fr12g-h,\,
G_+=g_+/h\equiv -\fr12G+1,\, G_-=g_-/h\equiv -\fr12G-1;
$
more detailed description can be found in [2].

In the timelike sector,
\be
g_-q<R^1<g_+q,
\ee
it is convenient to introduce the notation
\be
v=R^1/R^0\equiv v^1,\qquad u=\sqrt{(R^2)^2+(R^3)^2}\,/R^0\equiv v^{\perp}
\ee
together with
\be
M=\sqrt{1-u^2}
\ee
and use the functions
\be
Q(g;v,u)
=1-v^2-u^2-gvM
\equiv M^2-gvM-v^2>0
\ee
and
\be
V(g;v,u)
=\lf(v-g_-M\rg)^{G_+/2}\lf(g_+M-v\rg)^{-G_-/2},
\ee
so that
\be
F=R^0V.
\ee
We find
\be
\D{V}{v}=-(gM+v)\fr VQ,
\qquad
\D{V}{u}=-u\fr VQ.
\ee

In the FG-approach under development, we may (and shall)
trully follow the ordinary fundamental
special-relativistic
view,
which matches also the general-relativistic view,
 that the light front
is defined by the isotropic surface of the basic metric function
of the physical
space-time.
Namely,
if
in the traditional SR we take
\be
{\rm the\; pseudoEuclidean\; light\;  front\; equation:}
\quad
 S=0,
\ee
then under the Finslerian extensions
we should merely insert here
$F=F_{\{Finslerian\}}$
in place of
$S=F_{\{Riemannian\}}$
to get
\be
{\rm the\; Finslerian\; light\;  front\; equation:}
\quad
F=0.
\ee

Faced with feasibility of the FG-way of consistent treatment,
it is also worth adhering
straightforwardly to the view that the kinematic relativistic
transformations are of but
{\it a pure-passive meaning}.
Namely they merely specify
the variation rules for the vector components in going
from one RF to another RF, --
the four-dimensional vectors themselves remain unchanged, keeping their
directions in the
four-dimensional space-time.
Therefore,
to develop Finslerian kinematics
a researcher is invited to follow
\ses

\nin
{\bf The Tetrad-Kinematic Method}:
\be
F
\rightarrow
\{g_{pq}\}
\rightarrow
\{H^{(P)}_q\}
\rightarrow \quad
the \quad kinematic \quad transformations
\ee
which works well in purely-inductive way.
That is, given a FMF $F$, one should calculate
the associated metric tensor,
then find for the tensor the orthonormal tetrads,
and after that treat them as RSs for intertial observers to
deduce the precise
formulae for kinematic transformations (replicating in fact the method
applied ordinarily in tetradic topics of
 general relativity; see [3-5]).
For the relevant
 FMF (1.4),
 the method leads directly
to the extended kinematic transformations
 presented
explicitly in Section 3
(intermediate
calculations will not be shown, for they are closely similar to those
presented earlier
in [6-9]).

On so doing,
we apply direct calculations to arrive in Section 4 at
the important
conclusion
that,
for the Finslerian kinematic transformations obtained, the following
Proposition is valid.
\ses

\nin
{\bf PROPOSITION 1.1}. {\it The Transport Synchronization holds fine}.
\ses

\nin
Therefore, no conventionalistic ``vicious circularities"
would
enter the Finslerian
approach under development, for it is quite obvious that one can state
\ses

\nin
{\bf PROPOSITION 1.2}. {\it Given the light-front equation {\rm (1.16)} with a
particular metric function $F$. If the Transport Synchronization holds
as basic then
the Light Signal
Synchronization
Method, founded on Eq. {\rm (1.16)},
can be verified against it and thus
gets empirical, and vice versa.
}
\ses

\nin
The truth of these two Propositions, when taken in concert,
leads obviously to justifiable utilization of the
one-way light velocity.

The value of $g$ should thus
characterize the degree to which
Lorentz invariance is broken in  $U$.
The speed of light, of either  one-way type or round-trip type,
is no more a universal constant, being expected to be
anisotropic even in the input
rest frame $\Si$.

Indeed, by comparing (1.4) and (1.16) we get
\ses

\nin
{\bf PROPOSITION 1.3}. {\it The lightsurface in the input rest frame
$\Si$
is given by the equations
\be
\lf(
\sqrt{1+\fr14g^2}
+\fr12g\rg)^2(R^1)^2=
(R^0)^2-(R^2)^2-(R^3)^2, \quad if \quad R^1\ge 0,
\ee
and
\be
\lf(\sqrt{1+\fr14g^2}-\fr12g\rg)^2(R^1)^2=
(R^0)^2-(R^2)^2-(R^3)^2, \quad if \quad R^1\le 0.
\ee
}

\ses

In particular,
whenever $R^1=0$, propagation of light is going in
accord with the ordinary isotropic law:
\be
(R^0)^2-(R^2)^2-(R^3)^2 \quad for \quad the \quad
{\cal F\cal D-}
perpendicular
\quad directions,
\ee
so
$c^{\perp}=1$ holds true.

Alternatively, if propagation takes place along, \it resp. \rm
opposite to, the
$\cal F\cal D$-direction,
so that
$R^2=R^3=0$
and
$R^1>0$,
{\it resp.} $R^1<0$,
then for the
light velocity value
$c^+$,
\it resp. \rm
$c^-$,
we have
\be
c^+=
\sqrt{1+\fr14g^2}
-\fr12g,
\qquad
c^-=
\sqrt{1+\fr14g^2}
+\fr12g,
\ee
which entails
\be
\fr{
c^+-
c^-
}2
=-\fr12g
\ee
and
\be
\fr{c^++
c^-
}2
=
\sqrt{1+\fr14g^2}.
\ee

\ses
\ses
\ses

\ses

After which in Section 5 we
propose the Finslerian relativistic
extensions  for the
dispersion relations and for the four-momentum.
Basic conclusions will be summarized in the last Section 6.
\ses
\ses

\nin
\setcounter{sctn}{2}
\setcounter{equation}{0}

{\bf 2. ON INVOLVED PRINCIPLES AND IDEAS   }
\ses

{\bf Ghost of Synchrony}
has plagued the SR over a century.
The so-called ``clock synchronization problem" was given a strong
impetus by the
critical essays
 by Reichenbach [10], Gr{\"u}nbaum [11-12],
and
Jammer [13],
published to overcome ``insufficient attention to the conventions
hidden in procedures of
length measurements".
These authors
motivated the view that
no observational difference would arise if the one-way velocity
of light were assumed to be
anisotropic.

Indeed, when one simply raises the claim: ``Measuring light velocity implies
synchronizated clocks" $\cal V\cal S$
the opposed
claim: ``In order to synchronize clocks
one needs the one-way velocity of light",
the logical situation looks  {\it circular}.
An urgent analysis and discussian of the differing views of
synchronism
can be found in the recent survey [14] which has embraced
numerous  books and papers that bear
directly to the question.

However, the fact is that, though the synchrony-treatments
certainly refined many important aspects
of relativity,
circling various possible re-synchronization
methods is impotent to
lead to  a new
fundamental relativistic theory. For they merely redone clock
setups without affecting
the primary metric function, the latter was  fixed
to be of the
traditional
square-root type.
{\it No new relativistic physics is stemming from re-synchronization procedures}.

\ses

More than that,
in the Lorentzian SR
there exists
\ses

\nin
THE CANONICAL SYNCHRONIZATION
\ses

\nin
as presented by the Einstein synchronization
on the basis of
 the isotropic
and symmetric radar-way method
[15].
Apart that the method is highly practical, that  synchronization
in the Lorentzian SR
is
{\it equivalent to The Transport Synchronization}
provided that one follows the Tetrad Kinematic Method (1.17)
(which entails the
ordinary Lorentz transformations in case of the pseudo-Euclidean basic metric).

If one takes the  position that
``geometry is a convention",
than
slow clock transport method and
any context of
light signal prescriptions are logically independent of each other.
Under such a
free position, one remains uncertain of a physical way to avoid
conventionalistic
``vicious circularities"
and is doomed to make the one-way speed of light indeterminate.

Conversely, we can undermine that position to argue against
the logistics of
``the
geometrical conventionalism"
in simple terms: ``A physicist needs a particular metric function to be
a physicist", - and then to raise the simple question: "Would The Canonical
Synchronization remains being meaningful under a Finslerian metric extension?".

The true answer is ``Yes" in
view of Propositions 1.1 and 1.2 formulated
above
in Introduction.
The due method can be
properly-motivated
by geometrical
reasons.
\ses

\nin
{\bf Realism of one-way light velocity concept}
is seriously  supported by intuition of pragmatic-party physicists.
Indeed, though
\ses

\nin
QUESTION: Whether or not a one-way velocity of light is physically
meaningful quantity?
\ses

\nin
is deeply perplexing with
anarchy of possible re-synchronizations of distant clocks,
the photon is a
particle like other
quantized particles.
\ses

\nin
{\it
``After all light goes from a point to another in a well-defined way,
and it would be very strange if the true velocity were
forever inaccessible to us.''} [F.Selleri [16], p. 44]
\ses

\nin
In this connection, many keen and important
questions can be raised for. First of all,
we should ask: If one admits that an electron, when going from one
point to another, is given a certain physical velocity
in an ordinarily accepted sense, why a photon is not?
If a neutrino has a zero mass, it should be uncertain with the velocity
 like photons!
But if a neutrino has acquared any small rest mass,
it should have been discribed with a precise physically
measurable values of an operator of velocity?

Moreover, one is fraught with
\ses

\nin
COINING: A conventionality of
the
one-way
light velocity
{\it opposite to}
a reality of the
photon momentum.
\ses

\nin
Indeed, on the one hand, a
conventionality
is a strong
philosophical
and
logistic
argument.
However, if  the velocity of photon is
indeed
``a conventional entity'', then the momentum of photon should also be of some
discriminating nature.
On the other hand,
whenever  physicists evaluate relativistic-particle experiments,
they apply faithfully the conservation law of momentum,
particularly
for
the Compton Effect:
\be
p_{\ga}+p_{e}=p_{\ga}'+p_{e}',
\ee
where the photon momenta
$
\{p_{\ga},p_{\ga}'\}
$
are certain and measurable physical quantities
as well as the
electron ones
$
\{p_{e},p_{e}'\}.
$
If
the
one-way
light velocity
is
conventional,
why the momentum of photon is not?
{\it Eclecticism} of such a situation is apparent
from various physical standpoints, and any claim to favour
synchronism  {\it is coining}: the speed of light in one direction was never
measured in not a single experiment - but the photon
momenta were directly detected in
everyday established practice in numerous laboratories?

Alternatively, the right-sense realism
was assigned
to
the
one-way
light velocity
by C.M.Will [17]
{\it in terms of measurable quantities}: whenever the latters are properly considered to be
``true physical arguments of the light velocity'',
the experimental data are independent of the clock synchronization pocedures.
If one endeavours to analyze experiments on the basis of such a
clear physical
principle,
as was  done in [17],
the one-way
light velocity is fully
vindicated and
stops being an
``intelligible convention".

In any case, the Light Conventionality is deeply
connected with the question of Nature of Geometry.
The fact is that, traditionally,
the SR-works presupposed the square-root metric.
If, however, the question is investigated in the light of the FG-approach,
we may follow the realistic geometry-motivated way of reasoning actually in accord with the
famous  as well as pragmatic claims:
\ses

\nin
{\it``Laws of geometry and nature are complementary."}
(E.A.Milne [18], p. 17)
\ses

\nin
and
\ses

\nin
{\it``... one cannot maintain that distant simultaneity is conventional without
also maintaining that such basic quantities as the proper time metric
are conventional as well."} (M.Friedman [19], p. 77).
\ses

\nin
Indeed it is simply obvious that,
because The General Theoretical Relativistic
Framework indispensably involves  the  Light Behaviour  and the
Space-Time Metric as being important parts,
then ``testing one part means a test for the second part''.
That is to say, the two Parts {\it should go hand in hand} over the SR-physics!
\ses

\nin
{\bf Interferometer-type experiments of post-Lorentzian appeal}
can conventionally be devided in two following distinguished groups.
\ses

\nin
$I_1$:
{\it The turntable experiments}
have played a prominent role in the testing of light behaviour.
They include: first, the historically-valuable
Michelson-Morley type
experiments
(of which the most precise one
was carried out by Joos in 1930 [20])
used
 optical
interferometers
and,
 second,
the modern
highly-monochromatic
maser
experiments".

Proposals of respective maser-used techniques can be traced back to 1961:
\ses

\nin
{\it
``It can be hoped
because of the striking monochromaticity of the
radiation produced by infrared and optical masers that they will contribute
to a number of fundamental experiments
which
involve
very precise measurements or
 comparisons
of two lenghts or of two frequences.
One such example would be a highly improved experiment of the
Michelson-Morley
type."} [C.H.Townes, [21], p. 9].

\ses

This general program was specified as follows:
\ses

\nin
{\it
``If one maser
is rotated with respect to the other and changes in the relative frequency
of the two  measured, any change in the effective
optical path for the
two can be detected.
Thus one has the equivalent of a
Michelson-Morley
experiment."
} [{\it op. cit}, p. 11],
\ses

\nin
and
 was first designed in the ammonia-beam maser
{\it Cedarholm
and
Townes}
experiment [22-23],
-
with
two masers
had
 oppositely directed molecular beams
and
mounted on a ``rack
which could be rotated about a vertical axis".
The measurements
of changes in frequency upon rotation of
equipment
 through $180^o$ were
repeated
at intervals throughout a year
during
1958-1959.
The experiment was motivated by the nineteenth
century ether-drift concept
(and was based on the analysis of M\o ller [24]), so that in the
fraction $v/c$ the numerator was associated to
the earth's orbital velocity around
the Sun, and hence the estimation
$(v/c)^2\approx 10^{-8}$
was used.

The improved
{\it Brillet and Hall}
experiment
[25],
which
used two He-Ne highly
monochromatic masers mounted with axes perpendicular on a
turning
``granite slab"
to examine frequency changes upon the slab rotation
through $90^o$,
was reported in 1979.
Analytically,
 the Robertson's framework
[26]
was applied.
The instruments measured the relative change of
the optical path $ct=l$ between the two end
points of the intreferometer and
the  wave interference effect
was exploited.
The experiment is often highlighted as
``a more sensitive,
modern laser version of the Michelson-Morley experiment".
The sensitivity was reported as
$\De l/l=2.5\cdot 10^{-15}.$

In all the above experiments the zero changes in frequencies have been fixed.
The {\it Brillet and Hall} experiment
seems to have
closed
the epoch of turntable intereferometer experiments.
\ses

\nin
$I_2$:
{\it The monitoring-type experiments}
involve the old
ether-minded
{\it Kennedy and Thorndike experiment}
[27],
which
used
an unequal-arm
Michelson interferometer
and
a mercury
lamp,
and
the modern-type {\it Hill and Hall experiment}
[28]
which utilized instead two He-Ne lasers
together with computer store methods and started theoretically with the
so-called Mansouri-Sexl test theory
[29],
so
that
the numerator in the fraction $v/c$ was taken to be
the ($\approx 400$ km/s)-velocity
of the earth with respect to the
Cosmic
Microwave
Background.

The {\it Kennedy and Thorndike experiment}
reported in 1932
was differed from the Michelson-Morley experiment
in the significant respect
that the apparatus was fixed in an earth-grounded
 laboratory,
so
the idea was to observe the interference fringes
due to motion of earth over a period of months.

Similarly,
in
the {\it Hill and Hall experiment}
 the earth itself was regarded
as a``carrying platform",
if not as a ``rotating optical bench".
Siderial signals were searching for
during periods of 1986-1989 years.
In this ``laser-based
 Kennedy and Thorndike experiment"
the high
sensitivity of $2\cdot 10^{-13}$ for a siderial term
was gained.

The expectation in these experiments was
to
look for the variation
in the length
{\it
as the laboratory
apparatus is rotated in a
cosmic
space}.
Positive effects
were absent:
no fringe shifts due to either the diurnal or the seasonal
changes in the motion of the earth laboratory were observed.
\ses

\nin
{\bf Anisotropies of the earth's
local
scale}
are,
 obviously, unobservable in the $I_2$-type experiments
which are  entirely of
cosmos-born interest
 to trace possible
diurnal or siderial
effects
of
the earth's motion or rotation.
Oppositely, as instruments are actively  changing
local
orientation in the
earth-bound
 laboratory,
the outputs of the relevant
$I_1$-type
observations,
in which special rotational
platforms
are used
 to mount equipments on,
 may be sensitive to
search for such anisotropies.

Aimed to study the
latter possibility
on the basis of a convenient FG-metric, it is worth
applying the concept
of
$\cal F\cal D$-direction (introduced in Section 1),
which trace is a tiny deformation of the square-root metric.
In general, there may be examined
various
situations which don't comply with the spatial isotropy,
including
the following nearest possibilities.
\ses

\nin
$A_1$: {\it The earth-rotational anisotropic effects} might be expected
to meet in
a local
RF
$\Si$ firmly anchored to the earth;
\ses

\nin
$A_2$: {\it The radial-space anisotropies}
may be caused by
the dominating
spherical-mode distribution of matter inside the planet,
so the radial direction is a real
candidate
for the role of the
$
\cal F\cal D$-direction
and the terrestrial-lab interpretation of the preferred RF
$\Si$ is possible;
\ses

\nin
$A_3$: {\it The cosmos-conditioned anisotropies} may come to play,
as caused by the non-uniformity of
distribution of matter and energy in the embient space;
\ses

\nin
as well as the customary-sense case:
\ses

\nin
$A_4$: {\it The
Cosmic Microwave Background}
retains its relevance, provided that the experimental evidence
would make us
 to conclude that isotropy of the
Background
is not rigorously valid.
\ses

\nin
Of course,  there is no
{\it a priori}
reason to judge which of them is of
$\cal F\cal D$-nature
proper.
In each particular case one of the central experimental
questions is what is
``a degree of anisotropy'',
due answers thereto would involve estimating
the characteristic Finslerian parameter $g$
and thereby provide an experimental testing of the Lorentz transformations.
\ses

\nin
\setcounter{sctn}{3}
\setcounter{equation}{0}

{\bf\nin 3. EXPLICATED FINSLERIAN  KINEMATIC  TRANSFORMATIONS}
\ses

Using the convenient notation
(1.9)-(1.10)
and assuming, without any loss of generality, that
 motion is going in the plane $R^1\times R^2$, straightforward calculations (which intermediate steps were
presented in [2])
lead to the following generalized kinematic transformations:
\ses

\be
R^0=
\fr{1}{V(g;v,u)}
\Bigl[t+
(\fr vM+g)x+
\fr{u}{M}\sqrt{Q(g;v,u)}\,y\Bigr],
\ee
\ses
\ses
\ses
\be
R^1=
\fr{1}{V(g;v,u)}
(vt+
Mx)
,
\ee
\ses
\ses
\ses
\be
R^2=
\fr{1}{V(g;v,u)}
\Bigl[ut+
(\fr vM+g)
ux+
\fr{1}{M}\sqrt{Q(g;v,u)}\,y\Bigr],
\ee
\ses
\ses
\ses
\be
R^3
=
\fr{\sqrt{Q(g;v,u)}}{V(g;v,u)}z,
\ee
\ses

\nin
which inverse reads
\ses
\ses

\be
t=
\fr{V(g;v,u)}
{Q(g;v,u)}\Bigl[R^0-(v+gM)R^1-uR^2\Bigr],
\ee
\ses
\ses
\ses
\be
x=
\fr{V(g;v,u)}
{Q(g;v,u)}\Bigl[-\fr vMR^0+MR^1+\fr{vu}MR^2\Bigr],
\ee
\ses
\ses
\ses
\be
y=
\fr
{V(g;v,u)}
{\sqrt{Q(g;v,u)}}\fr1M(-uR^0+R^2),
\ee
\ses
\ses
\ses
\be
z=
\fr
{V(g;v,u)}
{\sqrt{Q(g;v,u)}}R^3.
\ee
\ses

\nin
Here, $Q$ and $V$ are the functions (1.11) and (1.12);
\be
\{R^0,R^1,R^2,R^3\}\in\Si, \qquad
\{t,x,y,z\}\in
S_{\{v,u\}},
\ee
where
$\Si$
is the input preferred rest frame and
$S_{\{v,u\}}$ is an inertial RF moving with the velocity
$\{v^1=v, v^2=u, v^3=0\}$ relative to
$\Si$;
the
instantaneously
common origin of the frames being implied.

Similarly for the momenta,
\ses

\be
P_0=
\fr{V(g;v,u)}
{Q(g;v,u)}
\Bigl[p_0-
\fr vMp_1-
\fr{u}{M}\sqrt{Q(g;v,u)}\,p_2\Bigr],
\ee
\ses
\ses
\ses
\be
P_1=
\fr{V(g;v,u)}
{Q(g;v,u)}
\Bigl[-(v+gM)p_0+
Mp_1
\Bigr],
\ee
\ses
\ses
\ses
\be
P_2=
\fr{V(g;v,u)}
{Q(g;v,u)}
\Bigl[-up_0+
\fr{ vu}Mp_1+
\fr{1}{M}\sqrt{Q(g;v,u)}\,p_2\Bigr],
\ee
\ses
\ses
\ses
\be
P_3=
\fr
{V(g;v,u)}
{\sqrt{Q(g;v,u)}}p_3,
\ee
and its inverse
\ses

\be
p_0=
\fr{1}{V(g;v,u)}
(P_0+
 vP_1+u
P_2),
\ee
\ses
\ses
\ses
\be
p_1=
\fr{1}{V(g;v,u)}
\Bigl[
(\fr vM+g)
P_0+
MP_1
+
(\fr vM+g)
u
P_2\Bigr],
\ee
\ses
\ses
\be
p_2=
\fr
{
\sqrt{Q(g;v,u)
}}
{V(g;v,u)}
\fr 1M
(uP_0+P_2)
\ee
\ses
\ses
\ses
\be
p_3
=
\fr{\sqrt{Q(g;v,u)}}{V(g;v,u)}P_3,
\ee
where
\ses

\be
\{P_0,P_1,P_2,P_3\}\in\Si, \qquad
\{p_0,p_1,p_2,p_3\}\in S_{\{v,u\}}.
\ee
\ses

\nin
The invariance of the contraction:
\be
R^0P_0+R^1P_1+R^2P_2+R^3P_3
=
tp_0+xp_1+yp_2+zp_3
\ee
can be verified directly.

Inversely, one may postulate (3.19) to explicate (3.10)-(3.17)
from (3.1)-(3.8).

$g$ is the characteristic Finslerian parameter, so that
the above transformations reduce exactly to the ordinary special-relativistic
Lorentz transformations whenever $g=0$.
\ses

\nin
\setcounter{sctn}{4}
\setcounter{equation}{0}

\nin
{\bf 4. TRANSPORT SYNCHRONIZATION}
\ses

If a clock moves in the RF
$S_{\{v,u\}}$
with a velocity
$\{\al,\beta,0\}$, so that
\be
\De x=\al \De t,
\qquad
\De y=\beta \De t,
\qquad
\De z=0,
\ee
we may consider the velocity
$\{w^1={\De R^1}/{\De R^0},\,w^2={\De R^2}/{\De R^0},\,0\}$
of the clock as viewed from the RF
$\Si$
and find from (3.1)-(3.3) the values
\ses
\ses
\ses
\be
w^1
=
\fr
{
v+
M\al
}
{
1+
(\fr vM+g)\al+
\fr{u}{M}\sqrt{Q(g;v,u)}\,\beta}
\ee
\ses
and
\be
w^2
=
\fr
{
u+
(\fr vM+g)u\al+
\fr{1}{M}\sqrt{Q(g;v,u)}\,\beta
}
{
1+
(\fr vM+g)\al+
\fr{u}{M}\sqrt{Q(g;v,u)}\,\beta}.
\ee
This entails
for the time
which the moving clock shows in its rest frame:
\be
 \De t'=\De R^0
V(g;w^1,w^2)\fr
{1-
(w^1)^2-(w^2)^2-gw^1
\sqrt{1-(w^2)^2}}
{
Q(g;w^1,w^2)}
\ee
The fraction here is unity (see (1.11)), so
\be
\fr{ \De t'}{\De R^0}=
V(g;w^1,w^2).
\ee
Applying here (3.1) yields
\be
\fr{ \De t'}{\De t}=
\fr{V(g;w^1,w^2)}
{V(g;v,u)}
\Bigl[1+
(\fr vM+g)\al+
\fr{u}{M}\sqrt{Q(g;v,u)}\,\beta\Bigr].
\ee

Let us follow the ideology of transport synchronization and consider
the clock motion to be infinitesimally slow with respect to the RF
$S_{\{v,u\}}$.
Then we approximate (4.2) and (4.3)  to first order
in $\al$ and $\beta$,
\be
w^1\approx
v+M\al
-v\lf(
(
\fr{v}{M}
+g)\al
+\fr uM\sqrt{Q(g;v,u)}
\,\beta\rg)
=v+\fr1MQ(g;v,u)\al
-\fr{ uv}M\sqrt{Q(g;v,u)}
\,\beta
\ee
and
\be
w^2\approx u+M
\sqrt{Q(g;v,u)}
\,\beta.
\ee
Since
\be
\D{V(g;v,u)}{v}=-(g
M
+v)\fr
{ V(g;v,u)}{Q(g;v,u)},
\qquad
\D{V(g;v,u)}{u}=-u\fr
{ V(g;v,u)}{Q(g;v,u)}
\ee
(see (1.14)),
we arrive at the vanishing
\be
\D{
(\De t'/\De t)
}{\al}_
{\Big|_{\al=\beta=0}}
=0,
\qquad
\D{
(\De t'/\De t)
}{\beta}_
{\Big|_{\al=\beta=0}}
=0,
\ee
which just
proves
\ses

\nin
{\bf PROPOSITION 4.1.} The Finslerian post-Lorentzian kinematic transformations
written out in Section 3
fullfill the Transport Synchrony.
\ses

\nin
\setcounter{sctn}{5}
\setcounter{equation}{0}
\ses

\nin
{\bf 5.  FINSLER-GENERALIZED HAMILTONIAN
AND
MOMENTUM}
\ses

Given
a particle of rest mass $m$.
For the Finslerian four-momentum
\be
\cP=\{P_0,P_1,P_2,P_3\}
\ee
of  the particle,
$P_0$
denotes the energy, $P_1$ is the projection of the four-momentum on the
preferred
$\cal  F\cal D$-direction, and
$
\{P_2,\, P_3\}$ is
the respective
perpendicular component.
It is convenient to put
\be
P_{\perp}=
\sqrt{(P_2)^2+(P_3)^2}.
\ee

In the time-like and isotropic cases of
the Finslerian approach under study, from (1.5) we
infer
{\it the Finslerian relativistic Hamiltonian}
\be
\cH(g;P_0,P_1,P_2,P_3)
=
\lf(-P_1+g_+\sqrt{(P_0)^2-(P_{\perp})^2}\rg)^{-G_-/2}
\lf(P_1-g_-\sqrt{(P_0)^2-(P_{\perp})^2}\rg)^{G_+/2}
\ee
which
is meaningful over the
 range of variation
\be
g_-\sqrt{(P_0)^2-(P_{\perp})^2}\,\le \, P_1\, \le \,
g_+\sqrt{(P_0)^2-(P_{\perp})^2},
\ee
where the equality relates to the isotropic case,
for which
$m=0$, and otherwise
the members
$\cP$
of the inequality (5.4)
are timelike.

So we get {\it the Finslerian dispersion relation}
\be
\cH(g;P_0,P_1,P_2,P_3)=m
\ee
to use it
in place of the ordinary pseudo-Euclidean
SR-prescription
$
\sqrt{(P_0)^2-(P_1)^2-(P_2)^2-(P_3)^2}
=m$.

The equation  (5.5) defines in an implicit way
the energy-momentum
function
\be
P_0=P_0
(g;m;P_1,P_{\perp})
\ee
which can be characterized by the derivatives
\be
\D{P_0}{P_1}=
\fr{P_1+g
\sqrt
{(P_0)^2-(P_{\perp})^2}
}
{P_0},
\qquad
\D{P_0}{P_{\perp}}=
\fr{P_{\perp}}
{P_0}
\ee
(obtainable on direct differentiating (5.5) for the case (5.3)).
In the slow-relativistic case, from (5.6)-(5.7) we get the approximation
\be
P_0\approx m+gP_1+\fr1{2m}
((P_1)^2+(P_{\perp})^2)+...
\ee
in which the first-order
Finslerian
 term,
$gP_1$,
has appeared.

In the Lorenzian SR-limit,
the definition (5.3) reduces to the ordinary pseudo-Euclidean one:
$$
\cH|_{_{g=0}}=
\lf(-P_1+\sqrt{(P_0)^2-(P_{\perp})^2}\rg)^{1/2}
\lf(P_1+\sqrt{(P_0)^2-(P_{\perp})^2}\rg)^{1/2}
=\sqrt{(P_0)^2-(P_1)^2-(P_{\perp})^2},
$$
that is,
\be
\cH|_{_{g=0}}=
\sqrt{(P_0)^2-(P_1)^2-(P_2)^2-(P_3)^2}.
\ee

We have
\ses

\nin
{\bf PROPOSITION 5.1.} \it In the Finslerian framework
under study, the occurence of the first-order term in
the slow-relativistic expansion \rm(5.8)
\it
of the energy-momentum dependence is characteristic\rm.
\ses

\nin
Indeed,
discarding such a term would entail
$g=0$ and, hence,
return us back to the ordinary pseuso-Euclidean relativistic dynamic relations.
\ses

\nin
\setcounter{sctn}{6}
\setcounter{equation}{0}

{\nin\bf 6. CONCLUSIONS}
\ses

Strictly speaking, the Finslerian relativistic framework
proposed is not
an implication of a free intelligible assumption about possibility of
generalization of the pseudo-Euclidean SR-theory, but rather is an accurate
{\it continuation} of the latter theory in
the Finslerian domain with respect to the single
 parameter, $g$.
Actually, the principal setups listed in Introduction allow the
relativistic FMF and the Finslerian kinematic transformations to be
introduced in a straightforward manner. The key parameter
$g$ appears on keeping the condition
that
the relativistic indicatrix,
as well as the mass shell
which is
given by (5.5),
be a space of constant negative curvature, which implies
for the curvature
$
R_{{\rm indicatrix}}=
R_{{\rm mass-shell}}=
-
(1+\fr{g^2}4)$
(see [2]).
By comparing the latter fact with (1.23),
we can conclude the remarkable relation
\be
\fr{c^++c^-}2=\sqrt{-
R_{{\rm mass-shell}}}\,.
\ee

A careful analysis  has shown that, as the nearest possibilities,
 there are two alternative ways for such
a
continuation.

The first way,
which was
developed in previous publications  [6-9],
 starts with the assumption
that the spatial isotropy is kept
valid
in a preferred rest frame.
The latter
assumption is
typically
accepted in  the
modern post-Lorentzian experimental as well as theoretical works
used
the Cosmic Microwave Background
 to be such a frame.
In this vein, a failure of Lorentz invariance would imply existence,
in any given terrestrial ground-based laboratory,
of a particular distinguished cosmos-born vector
which would
point the velocity of the earth with respect to
the Cosmic Microwave Background.
Experimental searches have been made for such a preferred RF in
various modern high-precision post-Lorentzian experiments, --
 which,
however,
have so far shown no evidence for such a vector.

In the second, alternative,
 way
 formulated in the present paper,
 the spatial anisotropy, rotational around a distinguished spatial direction,
is admitted
in
a preferred rest
frame, $\Si$,
so that the parameter  $g$  measures a
degree (intensity) of anisotropy.
The Principle of Correspondence is implied:
light signals are assumed to travel relative to $\Si$
with the common, isotropic and universal, speed $c_0$, unless FG-corrections
are switched on
(put $g=0$ in Eqs. (1.18) and (1.19)).
At low
velocities of relative motions, the Finslerian relativistic
relations differ by
but minor corrections from respective relations of ordinary SR-theory.

The Finslerian kinematic transformations obtained may serve
as providing various
attractive
direct
theoretical bases for post-Lorentzian estimations
in situations capacious of
revealing spatial isotropies,
and particularly for study the
local-space anisotropies of
``the earth's origin".
Although {\it a priori}
we may
have
little idea of the magnitude of the respective
characteristic Finslerian parameter $g$,
the developed Finslerian framework can well be adapted to the needs of relevant
post-Lorentzian SR-experiments when one is
 requested
 to create
self-consistent
relativistic
programs for testing local-scale, as well as cosmos-scale,
anisotropies.

Generally, the slow clock transport method can be viewed as allowing
a determination of
The Canonical  Synchronization
without any vicious circularity.
Having
the transport synchrony
proven
for the Finslerian kinematic transformations
under study
(Proposition 4.1),
we can undermine the pure-conventionalistic, or neetly-speculative,
 claim that
``the speed of light is illusion''.
The respective
anisotropy in the one-way
light velocity may be subjected to experimental search
(in particular, via measurements of due Finslerian corrections
to Doppler-effect phenomena), provided the FMF is fixed accordingly.
Simultaneously, experimental verification of the
Finslerian kinematic transformations which we have undertaken to propose
might show whether the primary geometry of space-time is Finslerian or not.

In the post-Lorentzian context
proper,
importance of ``the deep connection between kinematics and dynamics"
has been enphasized
in the highly readable
program by M.P.Haugan and C.M.Will [1].
Dynamically, the FG-approach is certain in that
all the basic dynamic relativistic ingredients are explicable
in a clear Finsler-geometrical way.
Indeed,
for the FMF
introduced by the definition (1.4),
the Finslerian Hamiltonian
(for a free relativistic particle)
is obtainable in an explicit form, presented by (5.3),
and then
the respective
Finslerian four-momentum is derivable directly to be
the momentum $\cP$ given by (5.1) subject to the Finslerian
dispersion relation
(5.5).
Therefore, a direct way is opening
to stretch  all the body of the Relativistic Dynamics in due
Finslerian way by using but a single characteristic parameter, g.
The principle of
momentum and energy conservation is maintained, but the equation relating energy to momentum
does change
in due Finslerian way (5.5).
In particular,
 for
the
Compton
effect
the respective Finslerian extension
of the
four-momentum conservation
law, as shown by (2.1),
can be written down in terms of the Finslerian
$\cP$-momenta.
We postpone generating
required
Finslerian analysis
 of particular physical
phenomena for further papers.

Obviously,
the theoretical structure and
experimental limitations of the special relativity
cannot be tested in all its aspects
unless
a self-consistent alternative theory
is applied.
We expect that it is the Finsler-type geometry that
may born such  theories,
advancing simultaneously {\it
new
philosophico-physical sights
on the relationship between Relativity and Geometry}.

An essential limitation of the $\cal F\cal D-$anisotropy
formulated in the present paper
is that
the anisotropy has been taken of ``uni-directional type", --
such that a preferred
vector of spatial type singles out one
distinguished direction.
Generally, anisotropies may be of complicated structure,
not predetermined by any single vector.
Of course, applying the uni-directional approach
is the easest way to think
of
Finslerian
correction to, or
violation
of,
the
pseudo-Euclidean
theory of
relativity.
\ses

\nin
\def\bibit[#1]#2\par{\rm\noindent
                     \parbox[t]{.05\textwidth}{\mbox{}\hfill[#1]}\hfill
                     \parbox[t]{.925\textwidth}{\bls11pt#2}\par}

{\nin\bf REFERENCES}\\
\ses

\nin
 1. M.P. Haugan and C.M. Will, \it Physics Today \bf 40\rm, 69 (1987).\\
 2. G.S. Asanov,  arXiv: gr-qc/0204070.\\
 3. J.L. Synge, \it Relativity: the General Theory
\rm (North-Holland, Amsterdam, 1960).\\
4. H.-J. Treder, H.-H. von Borzeszkowski, A. van der Merwe, and W~Yourgrau,\\
\it Fundamental Principles of General Relativity Theories. Local and Global
Aspects of\\
 Gravitation
and Cosmology \rm (Plenum, N.\,Y., 1980).\\
5. G.S. Asanov, \it Finsler Geometry, Relativity and Gauge
 Theories \rm (D.~Reidel Publ.
 Comp., Dordrecht, 1985).\\
6. G.S. Asanov, \it Rep. Math. Phys. \bf 39\rm, 69 (1997);
\bf 41\rm, 117 (1998);
\bf 46\rm, 383 (2000);\\
\bf 47\rm, 323 (2001).\\
7. G.S. Asanov, \it Moscow University Physics Bulletin
 \bf49 \rm (1), 18 (1994); \bf 51 \rm (1), 15 (1996);
\bf 51 \rm (2), 6 (1996); \bf 51 \rm (3), 1 (1996);
\bf 53 \rm (1), 15 (1998).\\
8. G.S. Asanov, ``The Finsler-type recasting of Lorentz transformations."
In: Proceedings
 of Conference  {\it Physical Interpretation of
Relativity Theory}, September 15-20 (London, Sunderland, 2000), pp. 1-24.\\
9. G.S. Asanov, \it Found. Phys. Lett. \bf15\rm, 199 (2002);
arXiv: gr-qc/0207089.\\
10. H. Reichenbach, \it The Philosophy of space and time \rm (Dover Pupl.,
Inc., N.\,Y., 1958).                            \\
11. A. Gr\"unbaum, \it The Philosophy of Science, \rm
A.~Danto and S.~Morgenbesser,
eds. (Meridian Books, N.\,Y., 1960).\\
12. A. Gr\"unbaum, \it The Philosophy of Space and Time
\rm (Redei, Dordrecht, 1973).         \\
13. M. Jammer,
``Some fundamental problems in the special theory of relativity", in
\it
Problems in the Foundations of Physics, \rm
G. Toraldo di Francia, ed. (Societa Intaliana di Fisica, Bologna, and
North-Holland, Amsterdam 1979), pp. 202-236.\\
14. R. Anderson, I. Vetharaniam and G.E. Stedman: \it  Phys. Reports
\bf 295\rm, 93-180
(1998).\\
15. A. Einstein, \it Ann. Physik \bf17\rm, 891 (1905).\\
16. F. Selleri,
\it Found. Phys. Lett. \bf 9\rm(1) (1996), 43.\\
17. C.M. Will, \it Phys. Rev. D \bf 45\rm, 403 (1992).                   \\
18. E.A. Milne, \it Kinematic Relativity, \rm (Oxford University Press,
Oxford, 1948).\\
19. M. Friedman, \it Foundations of Space-Time Theories
\rm
(Princeton University Press, Princeton, 1983).  \\
20. G. Joos, \it Ann. Physik \bf7\rm, 385 (1930).\\
21. C.H. Townes, ``Some applications of optical and infrared masers".
In: {\it Advances \\
in Quantum Electronics}
(Jay R. Singer, ed.)
(Columbia Univ. Press, N.Y. and London, 1961),
pp. 3-11.\\
22. J.P. Cedarholm and C.H. Townes,
\it Nature (London) \bf184\rm, 1350 (1959).\\
23. J.P. Cedarholm, F.G. Bland, B.L. Havens and C.H. Townes,
\it Phys. Rev. Lett. \bf1\rm, 526 (1958).\\
24. C. M{\o}ller, \it The Theory of Relativity\rm, 2nd edition
(Claredon, Oxford, 1972), p. 21. \\
25. A. Brillet and J.L. Hall, \it Phys. Rev. Lett. \bf 42\rm, 549 (1979).\\
26. H.P. Robertson: \it Rev. Mod. Phys. \bf 21\rm, 378 (1949).\\
27. R. J. Kennedy and E. M. Thorndike, \it Phys. Rev. B \bf 42\rm, 400 (1932).\\
28. D. Hils and J.L. Hall, \it Phys. Rev. Lett. \bf 64\rm, 1697 (1990).     \\
29. R. Mansouri and R. Sexl,
\it Gen. Rel. Grav. \bf8\rm, 496, 515, and 809 (1977) .\\

\end{document}